\documentclass[aps,twocolumn,showpacs,aps,groupedaddress,superscriptaddress]{revtex4}
\usepackage{graphicx}
\usepackage{dcolumn}
\usepackage{amsmath}
\usepackage{amssymb}

\begin{document}
\title{Charge and Orbital Ordering in Pr$_{0.5}$Ca$_{0.5}$MnO$_3$
Studied by $^{17}$O  NMR}

\author{A.~Yakubovskii}
\affiliation{Russian Research Centre "Kurchatov Institute",
Moscow,123182 Russia}
\affiliation{Laboratoire de Physique du
Solide, E.S.P.C.I., Paris, France}

\author{A.~Trokiner}
\affiliation{Laboratoire de Physique du Solide,  E.S.P.C.I.,
Paris, France}

\author{S.~Verkhovskii}
\affiliation{Institute of Metal Physics, Russian Academy of
Sciences,620219 Ekaterinburg, Russia}

\author{A.~Gerashenko}
\affiliation{Institute of Metal Physics, Russian Academy of
Sciences,620219 Ekaterinburg, Russia}

\author{D.~Khomskii}
\affiliation{Laboratory of Solid State Physics, Groningen
University, Netherlands}

\smallskip
\begin{abstract}
The charge and orbital ordering in Pr$_{0.5}$Ca$_{0.5}$MnO$_3$ is
studied for the first time by $^{17}$O  NMR. This local probe is
sensitive to spin, charge and orbital correlations. Two
transitions exist in this system: the charge and orbital ordering
at $T_{CO}=225K$ and the antiferromagnetic (AF) transition at $T_N
= 170K$. Both are clearly seen in the NMR spectra measured in a
magnetic field of $7T$. Above $T_{CO}$ there exists only one NMR
line with a large isotropic shift, whose temperature dependence is
in accordance with the presence of ferromagnetic (FM)
correlations. This line splits into two parts below $T_{CO}$,
which are attributed to different types of oxygen in the
charge/orbital ordered state. The interplay of FM and AF spin
correlations of Mn ions in the charge ordered state of
Pr$_{0.5}$Ca$_{0.5}$MnO$_3$ is considered in terms of the hole
hopping motion that is slowed down with decreasing temperature.
The developing fine structure of the spectra evidences, that there
still exist charge-disordered regions at $T_{CO}>T>T_N$ and that
the static ($t>10^{-6}s$) orbital order is established only on
approaching $T_N$. The CE-type magnetic correlations develop
gradually below $T_{CO}$, so that at first the AF correlations
between checkerboard ab-layers appear, and only at lower
temperature - CE correlations within  the ab-planes.
\end{abstract}


\maketitle
\smallskip
\section{Introduction}

The charge ordering phenomena in the hole-doped
R$_{1-x}$A$_x$MnO$_3$ (R is a trivalent rare-earth ion and A is a
divalent alkaline-earth ion) have been a subject of extensive
studies due to intriguing interplay of the charge, orbital and
spin degrees of freedom. The charge ordered (CO) state is formed
due to localization of the mobile $e_g$-holes.

Above $T_{CO}$  the $e_g$-holes provide FM correlations between
electron spins of neighboring manganese ions through the
double-exchange (DE) mechanism proposed by Zener
\cite{Zener-PRB82}. Pr$_{1-x}$Ca$_x$MnO$_3$ ($0.3 \leq x \leq
0.75$) oxides are the most suitable for investigation of the CO
state since the onsets of the charge ($T_{CO}$ ) and spin ($T_N$)
order are well separated in temperature. This doped oxide has an
orthorhombic structure (space group, $Pbnm$) in a wide temperature
and magnetic field ($<20T$) range. It remains in the
semiconducting state with no admixture of FM metal phase as
opposed to Pr$_{1-x}$Sr$_x$MnO$_3$, La$_{1-x}$Ca$_x$MnO$_3$ and
La$_{1-x}$Sr$_x$MnO$_3$ \cite{Jirak-JMMM53}, \cite{Tomioka-PRB53}.
In the CO structure of Pr$_{1-x}$Ca$_x$MnO$_3$ with $x \sim 0.5$
the in-plane pattern of Mn$^{3+}$ and Mn$^{4+}$ ions may be
represented as a checkerboard related to the corresponding
$t_g^3e_g^1$ and $t_g^3$ electronic configurations of Mn (Fig.1).
The lobes of a certain number of occupied $e_g$-orbitals are
ordered in the direction of the Mn$^{3+}$-O-Mn$^{4+}$ bond to
maximize DE coupling. Whereas the AF superexchange $t_{2g}-t_{2g}$
coupling is a dominating magnetic interaction for Mn$^{4+}$ and
Mn$^{3+}$ which $e_g$-lobes are aligned perpendicular to the
Mn$^{3+}$-O-Mn$^{4+}$ bond. At $x\sim 0.5$ the competition of
these exchange couplings results in the AF spin ordering of
CE-type with $T_N < T_{CO}$ . The associated Jahn-Teller (JT)
distortions of MnO$_6$ octahedra double the unit cell along b-axis
of the orthorhombic ($Pbnm$) lattice. The ideal CE type
charge/orbital order implies FM zigzag arrangement of the ordered
$e_g(3x^2-r^2)$ and $e_g(3y^2-r^2)$ orbitals of Mn$^{3+}$ ions in
ab-plane. The neighboring zigzags are AF-coupled, the ordering in
c-direction is also antiferromagnetic.

Recent resonant x-ray scattering \cite{Zimmerman-PRB61}, electron
microscopy \cite{Mori-PRB59} and neutron diffraction
\cite{Jirak-PRB61}, \cite{Kajimoto-PRB63} studies of
Pr$_{0.5}$Ca$_{0.5}$MnO$_3$  have shown that the orbital order
(OO) below $T_N=170K$ results in an orbital domain state
commensurate with the lattice. This commensurate OO becomes
metastable above $T_N$. Its melting is observed in diffraction
studies as a commensurate-incommensurate (C-IC) transition at
$T_{C-IC} \sim (180-200)K >T_N$. With further increase of
temperature the partial orbital disorder turns on the FM spin
fluctuations which become dominating near $T_{CO}\approx 250K$
\cite{Kajimoto-PRB58}. The C-IC transition was considered in terms
of the $e_g$-orbital polarization soft mode. The $e_g$-orbital
completely polarized along Mn-O bond corresponds to the amplitude
of the wave at a given Mn$^{3+}$ ion \cite{Jirak-PRB61} and its
wave vector $\textbf{q}=\{0, 1/2-\varepsilon, 0\}$ was considered
as the order parameter of the transition.

Further discussion of the CO state requires more detailed
microscopic data related to the distribution of spin density and
to its dynamic regime in the CO state of manganite. Studies of
spatial fluctuations of charge/orbital order with diffraction
experiments \cite{Jirak-PRB61}, \cite{Kajimoto-PRB63} is
restricted to short correlation times $\tau _c < 10^{-12}s$.
Whereas NMR experiments enable studying time-dependent spin
fluctuations at much longer $\tau_c$.

In this work we have studied the spin correlations of the
neighboring Mn ions developed in the paramagnetic charge ordered
(CO PM) state of Pr$_{0.5}$Ca$_{0.5}$MnO$_3$  by measuring the
$^{17}$O  NMR spectra. Oxygen atoms being placed between two Mn
ions bring valuable information about the spin/orbital
configuration of the nearest Mn ions. As expected the nuclear spin
of oxygen $(^{17}I=5/2)$ probes its magnetic state through the
dipolar and transferred hyperfine magnetic fields that depend on
the spin/orbital configuration of the neighboring Mn ions
\cite{Turov}. The $^{55}$Mn nucleus is a less suitable NMR probe
for this task for the following reasons. First, the NMR spectrum
of Mn$^{4+}$ ion ($t_g^3$ electron configuration) is available
only at low temperatures in the metastable CO AF phase. On the
other hand, NMR of Mn$^{3+}$($t_g^3e_g^1$) is hard to detect due
to extremely high nuclear spin-spin relaxation rate apparently
controlled by abnormal low-frequency spin dynamic of the localized
e$_g$-electrons.

This $^{17}$O  NMR report is focused on the study of the
development of the charge/orbital ordering in
Pr$_{0.5}$Ca$_{0.5}$MnO$_3$. The main result is that in
Pr$_{0.5}$Ca$_{0.5}$MnO$_3$  the CE-type magnetic correlations
develop gradually below $T_{CO}$: first those between ab-layers
arise and only at lower temperature the correlations within
ab-planes appear. The static ($\tau_c>10^{-6}s$) orbital order is
established only on approaching $T_N$.

\smallskip
\section{Experimental}

We used a powder sample of Pr$_{0.5}$Ca$_{0.5}$MnO$_3$  prepared
by traditional ceramic technology. The powder was enriched with
$^{17}$O isotope up to $\approx 25\%$. The single-phase nature of
the enriched sample was confirmed by the x-ray diffraction and
Raman scattering studies at room temperature.

$^{17}$O  NMR measurements were carried out on a phase-coherent
NMR pulse Bruker spectrometer over the temperature range of
$80-330K$ in magnetic field of $7T$. In this field the onset of
the CO and AF spin ordering was found to shift slightly down to
$T_{CO}\approx 225K$ and $T_N\approx 170K$ compared to reported
data at zero field and in accordance with $H$-dependence of
$T_{CO}$ reported in Ref.\cite{Tomioka-PRB53}. NMR spectra were
obtained for a loose-packed powder sample with point-by-point
frequency sweep measurements, the intensity of the spin-echo
signal formed with the pulse sequence
$(\pi/2)-\tau_{del}-(\pi/2)$-echo being measured. The width of the
$\pi/2$- exciting pulse was $t_p=(2-4.5)\mu s$ and the distance
between pulses varied in the range of $\tau_{del} = (40-80)\mu s$.
For each frequency, the amplitude of the exciting rf-pulse was
adjusted in order to optimize the echo signal intensity while
keeping the pulse duration fixed. All the echo-intensities were
corrected to $\tau _{del}\approx 0$ by measuring the rate of
echo-decay at different frequencies of the broad spectrum. All the
spectra for $T<T_{CO}$ were measured during cooling down from room
temperature to avoid the hysteresis uncertainties. The $^{17}$O
NMR signal in H$_2$O was used as a frequency reference to
determine the shift of NMR line in our sample.

\smallskip
\section{RESULTS AND DISCUSSION}
\subsection {The charge disordered paramagnetic state}

Figure 2 shows the $^{17}$O  NMR spectra of
Pr$_{0.5}$Ca$_{0.5}$MnO$_3$ measured in the PM state. In the
charge disorder paramagnetic state (CD PM) for $T>T_{CO}\approx
225K$ the main signal $(\sim 95\%)$ in  the spectrum is a rather
asymmetric line. It has a positive and extremely large magnetic
shift exceeding $6\%$ even at  the highest measured temperature,
$T=330K$ with respect  to those observed in nonmagnetic compounds.
An additional  line of small intensity $(\sim 5\%)$ is present at
zero NMR-shift.  It is supposed to originate from a small amount
of the Ca-based  oxides, which arise as spurious precipitations
during the  solid-state reaction synthesis. It's well known that a
small  concentration of spurious phase formed with light atoms
like  Ca is hard to detect by x-ray.

Let us now consider the quadrupolar and magnetic shift
interactions of the oxygen nuclear spin with its environment. The
oxygen atoms are located at the corners of octahedra with Mn at
the center. The noncubic local symmetry of oxygen sites leads to
the interaction of $^{17}$O  electric quadrupole moment $(eQ)$
with electric field gradient $(eV_{ZZ}$). The resulting $^{17}$O
NMR spectrum is expected to split into $2I+1$ lines separated by
the quadrupole frequency $\nu_Q=\frac{3e^2Q}{2I(2I-1)h V_{ZZ}}$ at
$\textbf{H} \parallel O_Z$ . Experimentally this characteristic
first-order quadrupole splitting has never been observed. It shows
that $^{17}\nu_Q$ does not exceed the NMR linewidth measured at
$T=330K$ i.e. $\nu_Q<1.4MHz$ in agreement with
Ref.\cite{Reven-Piskunov-Takigawa}. Thus the quadrupolar
interaction will only provide broadening effects on the observed
pattern of NMR spectra, these quadrupolar effects being small
compared to the magnetic interaction effects (see below).

The powder pattern spectrum of the Pr$_{0.5}$Ca$_{0.5}$MnO$_3$
main line in the CD PM state may be described assuming an axial
symmetry of the magnetic shift tensor
$\{K_\perp,K_\perp,K_\parallel\}$. The subscripts $"\perp"$ or
$"\parallel"$ refer to the shift component $K_\alpha$ of oxygen
with Mn-O bond directed perpendicular or parallel to the external
magnetic field \textbf{$H_0$}, respectively. The analysis of the
line leads to $K_\perp=7.0\%$ and $K_\parallel=11.5\%$ at $T =
295K$. The thermal variation of the shift is shown in Fig.3. With
decreasing temperature the peak of the line is further shifted
following the Curie-Weiss law with $K=K_0+a/(T -\theta)$. The
corresponding fit curve drawn by a dashed line in Figs. 2,3
results in $K_0=-1.0(6)\%$ and $\theta=130(20)K$. A close value of
$\theta\sim 150K$ is obtained by fitting the magnetic
susceptibility $(\chi)$ data measured in the range of $T=250\div
300K$ in Ref.\cite{Jirak-PRB61}. The positive value of $\theta$
evidences that FM spin correlations between neighboring Mn are the
dominating ones in the CD PM state above $T_{CO}$. A similar
conclusion about the prevalence of the FM spin correlations
between Mn ions in the CD PM phase was inferred from the
$^{139}$La NMR line shift in La$_{0.5}$Ca$_{0.5}$MnO$_3$, which
becomes FM-metal in the field of $4T$ below $T\approx 220K$
\cite{Yoshinari-PRB60}.

The slope of the $K$ vs $\chi$ plot corresponds to the local
magnetic field $H_{loc}=\mu_B\Delta{K}/\Delta\chi\approx 1\cdot
10^4 Oe/\mu_B$ (the $\chi$-data are taken from Fig.4 in
Ref.\cite{Jirak-PRB61}. It exceeds the magnitude of the classic
dipolar field $(H_{dip})$ induced at oxygen by the magnetic
moments of the neighboring Mn ions:
\begin{eqnarray}
H_{dip}=\frac {2g_e \mu_B\langle s_z(Mn)\rangle}
{r^3_{Mn-O}}=\frac{2 \chi H}{r^3_{Mn-O}}.
\end{eqnarray}
At room temperature $H_{dip}$ may be estimated as $H_{dip}\approx
 1200Oe$ where $\chi$ is defined per spin from Ref.\cite{Jirak-PRB61}.
The corresponding estimated anisotropic contribution to the total
NMR line shift $\{-0.8\%, -0.8\%, 1.6\%\}$ is much less than the
experimental value. Thus we assume that the classic dipole
interaction of $^{17}$O  nucleus with the effective magnetic
moments $2g_e\mu_B\langle{s_z(Mn)}\rangle$ of the nearest Mn has
no strong influence on the total shape of the spectrum.

The most probable origin of the observed giant isotropic magnetic
shift $K_{iso}\equiv(2K_\perp+K_\parallel)/3$ is the Fermi contact
interaction of the nuclear spin $^{17}I$ with the transferred
s-spin density of electrons participating both in the Mn-O-Mn
bonding and in superexchange coupling of the neighboring Mn,
\begin{eqnarray}
K_{iso}=\frac{8\pi}{3}g_e\mu_B\mid\phi_{2s}(0)\mid^2f_s\langle
{s_z(Mn)}\rangle=\nonumber\\
\frac{1}{^{17}\gamma\hbar}a(2s)f_s\langle{ s_z(Mn)}\rangle
\end{eqnarray}
Here $g_e \mu_B \langle{ s_z(Mn)}\rangle=\chi H$,
$a(2s)=^{17}\gamma\hbar{H_{FC}}(2s)=0.15cm^{-1}$ is the isotropic
hyperfine coupling constant for oxygen ion \cite{O'Reilly-JCP40},
\cite{Fraga-Handbook}. $H_{FC}$ is the corresponding hyperfine
magnetic field due to the Fermi-contact interaction with electron
located on 2s-orbital with wave function $\psi_{2s}(r)$. Following
\cite{Shulman-PR108}, \cite{Turov} the corresponding isotropic
spin density transferred at oxygen from neighboring Mn ions may be
defined in terms of the factor $f_s=H_{loc,iso}/2H_{FC}(2s)$. This
quantity estimated from (2) results in a rather large magnitude of
the effective fractional occupancy of O(2s) orbitals by unpaired
spins. We found $f_s=0.01$ for insulating PM state of
Pr$_{0.5}$Ca$_{0.5}$MnO$_3$. For comparison the LDA+U band
structure calculation results in $f_s=0.003-0.007$ for the
FM-ordered state of LaMnO$_3$ \cite{Elfimov}.

Another possible isotropic hyperfine interaction is the
core-polarization term which provides $H_{loc}\simeq 3 \cdot
10^4Oe$ \cite{Karplus} when assuming that the doped electron is
fully localized within the oxygen 2p-orbital. But this is not the
case above $T_{CO}$  in Pr$_{0.5}$Ca$_{0.5}$MnO$_3$. Thus this
interaction may be neglected for the light oxygen  atom.

The noncubic local symmetry of oxygen sites gives rise to an
anisotropic part of the magnetic shift tensor
$K_{ax}=(K_{\parallel}-K_{\perp})/3$. The most reasonable
contribution to $K_{ax}$ is the magnetic dipole-dipole hyperfine
interaction of $^{17}I$ with electrons on 2p$_{\sigma\pi,}$
orbitals, which spin occupancies 'up' and 'down' become different
through the polarizing interactions with neighboring manganese.
The corresponding magnetic shift tensor $K_{dip}$ may be expressed
through the fractional spin density transfer
$f_{\alpha}\langle{s_z(Mn)}\rangle$ on O-2p$_\alpha$ orbital from
paramagnetic neighboring Mn-ion \cite{Turov} as:
\begin{eqnarray}
K_{\parallel}(2p_\sigma)=-2K_{\perp}(2p_\sigma)=
\frac{4}{5}<r^{-3}>_{2p}f_\sigma \chi(Mn);
\end{eqnarray}
$K_\parallel(2p_\sigma)$,($K_\perp(2p_\sigma)$)  being defined for
\textbf{$H_0$} parallel (perpendicular) to the Mn-O bond
respectively. The resulting  dipolar contribution to $K_{ax}$ is
thus determined by the difference of transferred spin density
$(f_\sigma-f_\pi)\langle s_z(Mn) \rangle$ for different O-2p
orbitals from neighboring Mn-ion \cite{Turov}:
\begin{eqnarray}
K_{ax}=\frac{2}{5}<r^{-3}>_{2p}(f_\sigma-f_\pi)\chi(Mn)
\end{eqnarray}
Taking $<r^{-3}>_{2p}=4.97 a.u.$ \cite{Fraga-Handbook} for neutral
oxygen atom we obtain a rather large positive value
$(f_\sigma-f_\pi)\approx 0.04$. The direct observation of the
positive value of $(f_\sigma-f_\pi)$ proves for the first time
that the $p_\sigma$-orbital directed along the Mn-O bond is more
polarized than the two other $p_\pi$-orbitals.

We showed before that the $^{17}$O  NMR line position in the CD PM
state is mainly determined by the isotropic transferred hyperfine
coupling, arising from the hybridization of Mn(3d) and O(2s)
orbitals. In the ideal $Pbnm$ structure there is no overlap of
Mn($t_{2g}$)-O(2s) orbitals due to their orthogonality, while
$e_g$-orbitals of Mn have a rather strong overlap with s-orbital
of neighboring oxygen as shown in Fig.1. Hereafter we restrict
ourselves to the effects of overlap and covalency between
Mn($e_g$) and O(2s) orbitals.

\subsection{The charge ordered paramagnetic state}

Just below $T_{CO}$ the $^{17}$O  NMR spectrum is substantially
broadened. It splits into two parts as shown in Fig.2. The
low-frequency spectrum in the range of $K=(-2\div +10 \%$ is
asymmetric (A-line). Its width decreases gradually as the
temperature approaches $T_N$. The high-frequency spectrum is
approximately twice larger in intensity and forms a rather
complicated pattern which center of gravity is shifted to
extremely large positive $K>20\%$. At $T=170K\simeq T_N$ it splits
into two broad lines (B- and C-line) of about equal intensity and
peaked at $K_B\approx 40\%$ and $K_C\approx 55\%$, respectively.

According to x-ray and neutron diffraction studies
\cite{Jirak-PRB61}, the CO and OO of the CE-type becomes
commensurate in the CO PM phase only near $T_N$. If the domain
structure of the OO \cite{Zimmerman-PRB61},\cite{Jirak-PRB61} is
ignored, one finds four groups of the oxygen atoms differentiated
by the charge and/or by the direction of occupied $e_g$-orbital of
the nearest-neighboring Mn-ions. The first and the second groups
are formed by apical oxygen located between two Mn$^{3+}$-ions
(O1-site, shown in Fig.1) or between Mn$^{4+}$ (O2/O3-site),
respectively. Note that the $s_z$-projections of electron spins
are AF correlated for neighboring Mn from adjacent ab-planes. A
third group is formed by oxygen (O4) in the ab-plane which
participates in the AF coupling of neighboring Mn$^{4+}$,
Mn$^{3+}$ from adjacent zigzags. The last group (O5) is formed by
oxygen in the ab-plane located between two FM coupled Mn$^{4+}$and
Mn$^{3+}$ ions inside a zigzag. The concentration of each type of
oxygen atoms obeys the "structural" ratio $1:1:2:2$.

In order to assign A-C-lines measured in the CO PM phase we
analyze first the spectra near $T_N$ with resolved structure. We
compare the "structural" ratio with the experimental "NMR line
intensity" ratio, which is close to $2:2:2$. We also take into
account that the local magnetic fields at different oxygen sites
are formed through the Mn-O-Mn exchange interactions which are
short-range in space. In CO PM phase the effective magnetic moment
of Mn (shown in Fig.1 by arrows) is defined by its projection
$g_e\mu_B\langle{s_z}\rangle$ on the direction of the external
magnetic field. In turn it controls the sign of the corresponding
$^{17}$O NMR shift contributed by each of the two neighboring Mn.

A-line is attributed to the apical (O1 and O2/O3) sites whereas B-
and C-lines - to the in-plane ones. The magnetic shift of the
A-line is small, and the local field at the corresponding $^{17}$O
is comparable in magnitude to the classic dipolar field estimated
above (1). At these apical sites the hyperfine magnetic shifts (2-
4) are greatly reduced since both neighboring Mn-ions are in the
same valence state and their effective magnetic moments
$g_e\mu_B\langle{s_z}\rangle$ are AF correlated in the CO PM phase
of CE-type.

Let us now consider the high-frequency part of the spectra. C-line
demonstrating the largest positive shift may be attributed to
oxygen positioned in O5 sites whereas B-line is presumably due to
oxygen located in O4 sites. Indeed for oxygen in O5 site the
transferred s-wave spin density is maximal since within the zigzag
the lobe of the partially occupied $e_g(m_l=0)$ orbital of
Mn$^{3+}$ ion points toward the neighboring oxygen. Furthermore
the two neighboring Mn ions are FM correlated. For O4 we
definitely expect a rather large transferred hyperfine field for
the following reason. Although the spins of Mn in adjacent zigzags
are antiparallel, the O4 oxygen is "sandwiched" between
Mn$^{3+}$and Mn$^{4+}$ ions with different spin values and
different orbital occupations, i.e. with different covalency. So
that the transferred polarization from these two Mn ions should
not compensate as they do for the apical oxygen (O1, O2/O3). This
should again result in a substantial shift, although smaller than
for O5. Moreover the transferred s-wave polarization from
Mn$^{4+}$ ion is expected to be negative due to effects of
covalent mixing with the empty $e_g$-orbitals \cite{Watson-PR134}.
As reviewed in \cite{Watson-HI} the charge transfer from the
occupied O-2s orbital to the empty eg-orbital is spin dependent.
It is regulated by the intra-atomic exchange coupling with
electrons on $t_g$-orbitals.

Thus in the CO PM phase the static s-wave polarization is directed
along $\textbf{H}$, and the isotropic shift at O5 site may be
expressed similarly to (2) through the corresponding transferred
spin densities $f_{s,3+}\langle{s_z(Mn^{3+}})\rangle$ or
$f_{s,4+}\langle{s_z(Mn^{4+}})\rangle$ of the neighboring Mn (with
$f_{s,3+}>0$ and $f_{s,4+}<0$):
\begin{eqnarray}
  K_{iso}(O5)=\frac{8\pi}{3}g_e\mu_B\mid\psi_{2s}(0)
  \mid^2\{f_{s,3+}\langle{s_z(Mn^{3+}})\rangle+\nonumber\\
  f_{s,4+}\langle{s_z(Mn^{4+}})\rangle\}
\end{eqnarray}
The effective magnetic moments of neighboring Mn from adjacent
zigzags are AF correlated and we obtain for $K_{iso}$(04) the
following expression:
\begin{eqnarray}
  K_{iso}(O4)=\frac{8\pi}{3}g_e\mu_B
  \mid\psi_{2s}(0)\mid^2\{0.25f_{s,3+}\langle{s_z(Mn^{3+}})\rangle-\nonumber\\
  f_{s,4+}\langle{s_z(Mn^{4+}})\rangle\}
\end{eqnarray}
The sign $"+"/"-"$ in (5),(6) takes into account the fact that
FM/AF spin correlations of the neighboring Mn$^{3+}$ and Mn$^{4+}$
ions are considered as static in the time interval which is much
longer that the inverse splitting of the B- and C-line
($\sim10^{-8}s$) of the spectrum at $T=170K\approx T_N$, when the
CO and OO are completely formed in PM state. This agrees with the
fact that the O(2s)-polarization is provided predominantly by the
$\sigma$-overlap with the $e_g$-orbitals of Mn, and Pauli blocking
of part of $e_g$-orbitals in Mn$^{3+}$ gives stronger polarization
than in the case of Mn$^{4+}$,where both 2s$\uparrow$ and
2s$\downarrow$-electrons of O2- may virtually hop to Mn$^{4+}$ so
that the net oxygen polarization due to covalency with Mn$^{4+}$
will be smaller than for Mn$^{3+}$-neighbor and negative. For
crude estimate of the transferred s-spin density we have assumed
in (5),(6) that the $\sigma$-overlap of $e_g(m_l=0)$ orbital of
Mn$^{3+}$ with 2s-orbital of O5 in zigzag of the CE-type charge
and orbital ordered phase is twice larger than the overlap with
corresponding 2s-orbital of oxygen O4 located between the
neighboring zigzags in ab-plane. (\textit{Of course under detailed
consideration it should depend on interatomic distance and on the
bending of the Mn$^{3+}$-O bond in the tilted and JT distorted
MnO$_6$ octahedra.}) Inserting the value of the peak position of
B-line $(K_{iso}=40\%)$ and C-line $(K_{iso}=55\%)$ into
expressions (6),(5) respectively we get that the transferred
s-wave polarization from Mn$^{3+}$ ion is positive and its
absolute value exceeds about 4 times the corresponding negative
polarization transferred from Mn$^{4+}$
$\mid{f_{s,4+}}\langle{s_z}(Mn^{4+})\rangle\mid$.

The large difference in spin densities transferred from
Mn$^{3+}$/Mn$^{4+}$ ions indicates a substantial delocalization of
the "$e_g$-hole" within the hybridized
$e_g(Mn^{3+})$-2p$_\sigma$(O) orbital \cite{Anisimov-PRB55}. It
shows that a pure ionic approach where the $e_g$-hole is
completely localized at the Mn$^{3+}$ ion is a very rough
approximation to describe in detail the CO and OO in doped
manganite.

Unfortunately the spatial distribution of the Fermi contact
hyperfine fields results in a large broadening of the separate NMR
lines in spectrum measured at $T<T_{CO}$. It masks the anisotropy
of the magnetic shift tensor and does not permit to trace the
transfer of the 2p-spin density at oxygen in the CO PM phase, i.e.
to address the orbital order more directly by studying
polarization of a given p-orbital.

\subsection{The antiferromagnetic state}

Different local fields for apical oxygen in O1 and (O2/O3) sites
are expected only in the spin-ordered phase below $T_N$. In the AF
phase the AF moment canting takes place when a rather high
magnetic field ($7T$) is applied. The canted moment of Mn in
adjacent ab-planes will create at the apical oxygen an additional
local field dependent on the valence state of neighboring Mn. At
O1 sites the neighboring Mn$^{3+}$ ions create a larger isotropic
magnetic shift than Mn$^{4+}$ ions do at O2/O3 sites. Indeed, as
shown in Fig.4 at $T=100K$ A-line splits into two lines of roughly
equal intensity. A loose packed powder sample in a strong enough
external field may be considered as partially oriented with
\textbf{c$\parallel$H}. It should be noted that the observed
splitting (~4MHz) is not a result of the classic dipolar fields of
the neighboring Mn ions since the difference in their effective
magnetic moments is too small ($\mu_{eff}=2.7\mu_B$ for Mn$^{3+}$
and $2.2\mu_B$ for Mn$^{4+}$ as estimated from Fig.4 in
Ref.\cite{Jirak-PRB61}).

The only slight additional broadening of the NMR spectra even
around $T_N/2$ evidences that the line shift is mainly determined
by the short-range CO and OO which have been completely formed in
the CO PM phase at $T\rightarrow{T_N}$. Below 110K it appears that
the optimal rf pulse duration increases about twice from the low
to the high frequency part of the A-line as illustrated in Fig.5.
The same feature was found for the high-frequency spectrum when
the B- and C-lines are resolved. This striking effect of the
nonmagnetic ligand atom on the echo formation may be related to a
rather strong isotropic hyperfine interaction between $^{17}$O
nuclear spin and the Mn electron spin system with long-range AF
spin order. The detailed analysis of the spectra measured below
$T_N$ requires additional studies which are now in progress.

\subsection{Development of charge and orbital ordering}

Based on the site assignment considered above we propose the
following picture of the development of spin correlation in the
Pr$_{0.5}$Ca$_{0.5}$MnO$_3$  CO PM phase as seen by $^{17}$O  NMR.
In the low-temperature part of the CO PM phase the NMR spectrum
represents the spin density distribution on oxygen ions. Its value
is determined not just by the Mn effective magnetic moment but, to
a greater extent, by the type of spin correlations of the
neighboring Mn atoms. The presence of several lines in NMR
spectrum, which can be attributed to the various oxygen sites in
the lattice, shows that the corresponding spin correlations and
effective magnetic moment of the neighboring cations do not change
at the time scale $t\geq 10/(\Delta \omega)\approx 10^{-6}s
(\Delta \omega$ - line splitting) \cite{Abragam}. The thermally
activated hopping of $e_g$-holes seems to be the main mechanism
changing the charge state of the ion (Mn$^{3+}$-Mn$^{4+}$) and the
spin correlations between the neighboring ions, which results in
the "melting" of OO in CO PM phase. With increasing temperature
the fine structure of the spectrum is smeared as the corresponding
correlation time $(\tau_c)$ of the specific hoping becomes
comparable with $(\Delta \omega)^{-1}$. As a result the various
spin configurations are no more distinguishable at higher
temperature.

The splitting of the NMR spectrum just below $T_{CO}$ into the A-
and (B+C)-lines  may be explained as follows. As has been
mentioned above the oxygen nuclei responsible for A-line are
located between two AF correlated Mn ions from adjacent ab-plane
whereas those responsible for (B+C)-line have two neighboring Mn
ions in the same ab-plane. The resolution of A-line shows that
just below $T_{CO}$  the correlation time of the AF correlated
spin of neighboring Mn from adjacent ab-planes becomes long
compared to the NMR time scale. By contrast (B+C)-line is still
unresolved just below $T_{CO}$ . Thus in the first stage of the
charge ordering the three-dimensional (3D) motion of eg-electrons
transforms preferentially into a two-dimensional (2D) hopping
within the ab-planes. Such an ordering phase transition may be
considered as the nucleation of AF ordered clusters which grow at
the expense of the CD PM phase which dominates above $T_{CO}$.
Furthermore as the temperature decreases and the unresolved
(B+C)-line is shifted toward high frequency a shoulder appears on
the low frequency part (see spectra for $T\approx 200K$ at Fig.
2). The temperature dependence of this signal follows the same
Curie-Weiss law as the peak in the CD PM phase (dashed line in
Figs. 2, 3). Its relative intensity decreases with temperature and
becomes negligible only near $T_N$. This may indicate that almost
down to $T_N$  traces of the CD PM phase remain in the CO PM
phase.

The proposed NMR interpretation of the ordering in the CO PM phase
is in a good agreement with the main results of the resonant x-ray
\cite{Zimmerman-PRB61} and neutron diffraction studies
\cite{Jirak-PRB61}, \cite{Kajimoto-PRB63}, \cite{Kajimoto-PRB58}.
It should be noted that the time scales required to get a
quasi-static picture of charge distribution in NMR
$(t_{nmr}>10^{-6}s)$ and in neutron diffraction $(t_{nd} \sim
10^{-12}s)$ experiments are very different. Both methods confirm
the presence of FM correlations between Mn as dominant spin
fluctuations in the CD PM phase which reduce in intensity below
$T_{CO}$ . In the neutron scattering experiments the short-range
FM correlations are considered as a static whereas it is seen
still as a low-frequency dynamic phenomenon in the NMR spectra. In
conclusion, the distribution of spin density and the development
of the charge and orbital ordering in the paramagnetic state of
Pr$_{0.5}$Ca$_{0.5}$MnO$_3$ were studied for the first time by
means of $^{17}$O  NMR. It is shown that the main interaction of
oxygen is the isotropic hyperfine Fermi-contact interaction with
the s-electron spin density transferred from the
nearest-neighboring Mn ions. The resulting magnetic line shift is
very sensitive to both the electronic configuration of the
neighboring cation (Mn$^{+3}$/Mn$^{4+}$) and to their specific
spin-pair correlations that develop on cooling in the charge and
orbital ordered PM phase. It is shown that with increasing
temperature the melting of the orbital ordering first develops
within the ab-plane whereas the AF correlations between Mn ions in
adjacent layers are more stable and disappear only when T
approaches $T_{CO}$ .

\begin{acknowledgments}
We are very grateful to Prof.A.~Kaul for supplying us the starting
Pr$_{0.5}$Ca$_{0.5}$MnO$_3$  material and to Dr.A.~Inyushkin for
$^{17}$O  isotope enrichment. The work is supported partly by
Russian Foundation for Basic Research (Grant 02-02-16357a) as well
as by CRDF RP2-2355 and INTAS 01-2008 Grants. We are especially
grateful for support of A.Y. by ESPCI.
\end{acknowledgments}

\newpage

\begin{figure}[!]
\includegraphics[width=0.45\textwidth]{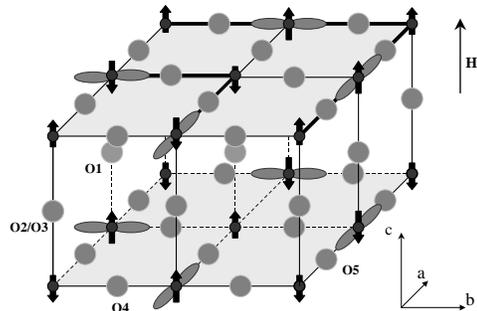}
\caption{Charge and orbital-ordered Pr$_{0.5}$Ca$_{0.5}$MnO$_3$
with CE-type magnetic structure:
 $\bullet$  - Mn$^{4+}$-ions;
 $\circ\bullet\circ$ - Mn$^{3+}$-ions with occupied $e_g$-orbitals;
 $\circ$  - oxygen ions (O1 -O5).
 The CE-type of spin correlations in the charge ordered
paramagnetic phase is shown by arrows directed up/down relative to
the applied magnetic field \textbf{H$\parallel$c}} \label{fig1}
\end{figure}

\begin{figure}[!]
\includegraphics[width=0.45\textwidth]{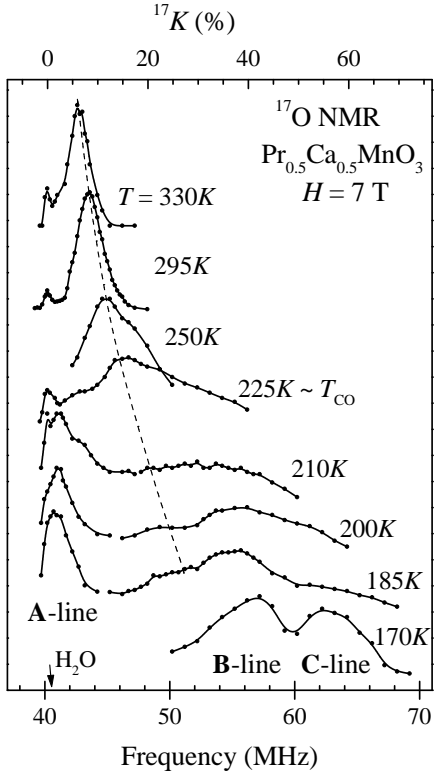}
\caption{$^{17}$O NMR spectra measured at $H=7T$ in the
paramagnetic state of Pr$_{0.5}$Ca$_{0.5}$MnO$_3$. The dashed line
represents the fitting curve of the line peak shift by the
expression $K_0+a/(T-\theta)$ for spectra measured above
$T_{CO}$.}
\end{figure}

\begin{figure}[tl]
\includegraphics[width=0.45\textwidth]{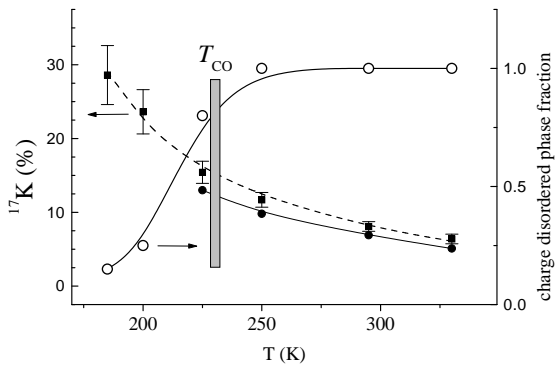}
\caption{ a) The line peak shift $K$(solid squares) $^{17}K_\perp$
(solid circles) vs.T plot. The dashed curve is the fit of $K$ data
by the expression $K_0+a/(T-\theta)$ with $K_0=-1.0(6)\%$ and
$\theta=130(10)K$ for spectra measured above $T_{CO}$; solid line
represents the fitting curve of the line peak shift by the
expression $K_0+a/(T-\theta)$; b)Relative $^{17}$O NMR line
intensity of oxygens in the charge disordered regions of PM
phase.}
\end{figure}

\begin{figure}[t]
\includegraphics[width=0.37\textwidth]{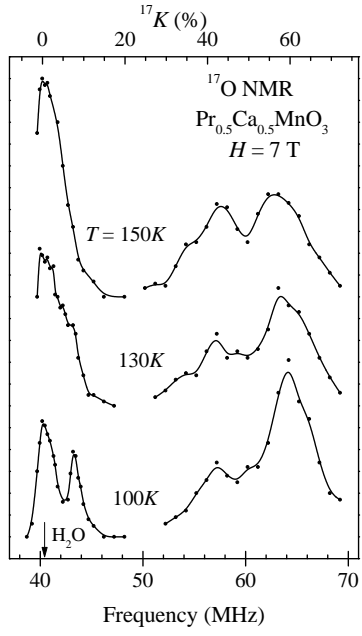}
\caption{ $^{17}O$ NMR spectra measured at $H_{ext}=7T$ in the AF
phase of Pr$_{0.5}$Ca$_{0.5}$MnO$_3$.}
\end{figure}

\begin{figure}[b]
\includegraphics[width=0.37\textwidth]{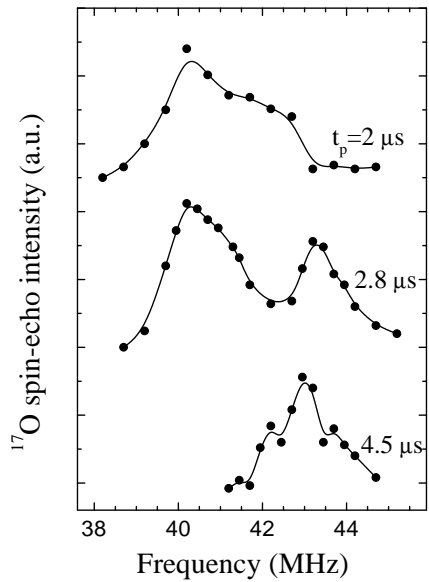}
\caption{Evolution of the echo-spectra shape of A-line measured
with different rf-pulse durations $(t_p)$ at $T=100K$ in the AF
phase.}
\end{figure}

\end{document}